\documentclass[twocolumn,prb,showpacs,preprintnumbers,amsmath,amssymb]{revtex4}

\usepackage{graphicx}
\usepackage{dcolumn}
\usepackage{bm}
\usepackage{bbm}
\usepackage{pstricks}

\newcommand{\llangle}{\mathop{\langle\negthinspace\langle}}
\newcommand{\rrangle}{\mathop{\rangle\negthinspace\rangle}}
\newcommand{\new}[2]{#2}

\begin{document}


\title{Weak coupling approximations in non-Markovian Transport}

\author{Philipp Zedler}
\email{zedler@physik.tu-berlin.de}
\author{Gernot Schaller}%
\author{Gerold Kiesslich}%
\author{Clive Emary}%
\author{Tobias Brandes}%
\affiliation{%
Technische Universit\"at Berlin, Hardenbergstr.\ 36, 10623 Berlin
}%


\date{\today}

\begin{abstract}
We study the transport properties of the Fano-Anderson model with
\new{1}{non-Markovian effects, which are introduced by making one tunneling rate
	energy-dependent. We show}
that the non-Markovian master equation
may fail if these effects are strong.
We evaluate the stationary current, the zero frequency current noise
and the occupation dynamics of the resonant level  by means of a quantum master equation approach  within different
approximation schemes and compare the results to the exact solution
obtained by scattering theory and Green's functions.
\end{abstract}

\maketitle

\section{Introduction}
The ongoing progress in measuring tiny fluctuations of charge currents
through nanoscale conductors \cite{BLA00,NAZ03,BLA05} 
has led to an increased theoretical interest in
non-Markovian effects revealed in such experiments \cite{KIE07b,BRA08a}. 
Of particular interest are the non-Markovian dynamics induced by the coupling to fermionic \cite{BRA06,KNE08} and/or bosonic \cite{AGU04,BRA05,CHE08} environments and their influence on 
steady-state transport observables such as the cumulants of the stochastic charge transfer process \cite{FLI08}. 
Our present work focuses on such effects due to the coupling to electronic reservoirs.

There exist various techniques for describing
open quantum systems coupled to fermionic reservoirs:
e.g., scattering theory and Green's functions 
\cite{DAT95,DAT05,Haug08},
or quantum master equations starting from the von-Neumann equation for the total density operator \cite{Breuer06}, 
\new{2}{
	or the Wigner-Boltzmann approach
	\cite{Nedjalkov04,Nedjalkov06,Querlioz08}.
	Master equations are
}
widely considered with
 a conductor-lead coupling in Born-Markov approximation (see e.g. \cite{STO96,GUR96c,KAI06}),
 which is only \new{9c}{strictly} valid for weak coupling and constant contact density of states in the energy range of interest.
 Consequently, to explore effects beyond the common \new{9c}{Born-}Markov-approximation,
 one would study higher-order perturbation theory in the contact coupling
 (e.g. \cite{THI04a,LI05,PED05}) and/or allow
 for energy-dependent tunneling rates, i.e., go beyond the wide-band approximation
 \cite{Haug08,Elattari00a,Matisse08,Thorwart04,Huang08,Kleinekathoefer04}.
In this work we choose the latter whilst analyzing
perturbative approaches in lowest-order tunnel coupling only.
In particular, we address the question of in which limits
it is possible to describe non-Markovian physics with master equations.

A conceptionally simple  model for this purpose
consists of a single resonant level
(e.g. the ground state of a quantum dot or a molecule)
weakly coupled to two electronic leads in equilibrium
and a Lorentzian-shaped density of states for one of the leads.
This model is equivalent to two serially
coupled quantum dots \cite{BRA05}
and can be interpreted as a quantum dot coupled to a reservoir
with finite electron relaxation time.

To enable an exact solution, and for the sake of simplicity,
we neglect Coulomb interaction and, consequently,
prominent effects like Coulomb blockade or Kondo correlations.
This, of course, constrains the use of our results
for quantitative understandings of transport experiments.
Our aim, however, is a comparison
of different approximation schemes for master equations
in an electronic transport problem that has an exact solution.

We will show that the non-Markovian master equation (NMME) in the wide-band
limit produces reasonable results for the current and noise.
However, reducing the band-width yields qualitative and quantitative
deviations --- even the emergence of unphysical results, like negative Fano factors.
\new{6 and 11}{}
The exact time evolution of the resonant level occupation can only be
obtained by the NMME in the wide-band limit
where even a Markovian master equation covers the exact dynamics.
In the short-time limit, the NMME result well approximates the
exact evolution regardless of the bandwith.
However, for very small bandwidths, the NMME generates negative, unphysical
probabilities. We demonstrate how this can be avoided by a dynamical coarse
graining method.

The paper is organized as follows:
In Secs.~II.A. and II.B. we introduce the model and provide
the known exact solution obtained by scattering theory and Green's functions.
In Sec.~II.C.
the equivalence to the double-dot model is discussed.
In Sec.~III.A. we introduce the non-Markovian master equation and in
Sec.~III.B. the dynamical-coarse-graining approach.
In Sec.~IV the steady-state current and the Fano factor are compared
to the exact and NMME solution.
Finally, the occupation dynamics of the resonant level is discussed.

\section{Model}

\subsection{Hamiltonian}
\begin{figure}
\includegraphics[width=.7\linewidth]{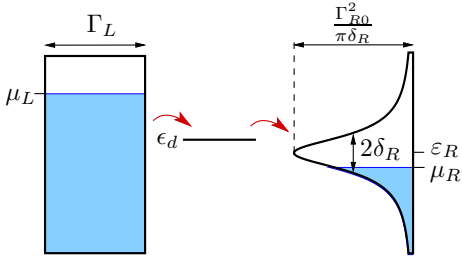}
\caption{\label{Model}
	Single resonant level coupled to two leads with one constant
	and one Lorentzian tunneling rate.
	In the infinite bias limit, \mbox{$\mu_L\to\infty$} and \mbox{$\mu_R\to-\infty$}.}
\end{figure}
We start from the well-known single resonant level model
\cite{Anderson61,Fano61,Mahan00, Haug08, Bruus04, Hewson93}
that is described by a Hamiltonian where tunneling between
two leads can be realized via a localized
quantum dot state,
\begin{equation}
H = \epsilon_d d^\dagger d
  + \sum_{k,a} \epsilon_{ka} c^\dagger_{ka} c_{ka}
  + \sum_{k,a} \left( t_{ka} d^\dagger c_{ka} + \mathrm{h.c.} \right).
\label{Hamiltonian}
\end{equation}
The fermionic operators $d$/$d^\dagger$ annihilate/create an electron 
on the dot and $c_{ka}$/$c_{ka}^\dagger$ annihilate/create an electron
in lead $a\in\{R, L\}$ with momentum $k$.
We choose a Lorentzian-shaped right tunneling rate, like in \cite{Elattari00a,Matisse08}:
\begin{equation}
\Gamma_R(\omega)
  = \frac{\Gamma_{R,0}^2}{\pi}
    \frac{\delta_R}{(\omega-\varepsilon_R)^2+\delta_R^2},
\label{Gamma}
\end{equation}
and a flat left tunneling rate $\Gamma_L$,
both related to the microscopic parameters via
\mbox{$\Gamma_a(\omega)=2\pi \sum_{k}|t_{ka}|^2\delta(\omega-\epsilon_{ka})$}.
A sketch of the model is given in Fig. \ref{Model}.
We work in the infinite bias limit, where we
couple the left and right lead to two particle reservoirs
in a way that the left lead is completely occupied
	and the right lead is completely empty, i.e.,
\mbox{$\mu_L\to\infty$} and \mbox{$\mu_R\to-\infty$}
such that the left Fermi function takes the value $1$
and the right one takes the value $0$.

The retarded and advanced self energies of our model
are evaluated exactly \cite{Haug08,Mahan00,Hewson93} as
\begin{equation}
\Sigma^{R/A}(\omega)
  = \sum_{k,a}\frac{|t_{ka}|^2}{\omega-\epsilon_{ka} \pm i0^+} = \Lambda(\omega) \mp i\frac{\Gamma(\omega)}{2}.
\label{SelfEnergy}
\end{equation}
The imaginary part
$\Gamma(\omega)=\Gamma_L(\omega)+\Gamma_R(\omega)$
is connected to the inverse lifetime,
\new{10}{
whereas the real part
\begin{equation}
\Lambda(\omega) = \frac{1}{2\pi}\mathcal{P}\int\limits_{-\infty}^\infty d\omega'
  \frac{\Gamma(\omega')}{\omega-\omega'}
  \label{KramersKronig}
\end{equation}
(where $\mathcal{P}$ denotes the principal value)
induces a level shift which
is fully determined by the tunneling rate (Kramers-Kronig relation).}
It will turn out that all quantities concerning
transport statistics and the occupation of the quantum dot
can be expressed in terms of
$\Gamma_a(\omega)$ and $\epsilon_d$ only.

\subsection{Exact solution}
To evaluate transport quantities we need the 
transmission coefficient \cite{Stone85},
\begin{equation}
T(\omega)
=	\frac{\Gamma_L(\omega)\Gamma_R(\omega)}
	  {|\omega-\epsilon_d-\Sigma^R(\omega)|^2},
	\label{SingleDotT}	\\
\end{equation}
which is related to the dot spectral function
\mbox{$A(\omega):= i\left[G^R(\omega)-G^A(\omega)\right]$} via
\mbox{$T(\omega)=\frac{\Gamma_L(\omega)\Gamma_R(\omega)}{\Gamma(\omega)}A(\omega)$}.
Scattering theory \cite{BLA00}
then yields the current $I$ and the zero-frequency noise $S$ which at infinite bias are
\begin{eqnarray}
I & = &	e\int\limits_{-\infty}^\infty\frac{d\omega}{2\pi} T(\omega),
	\nonumber	\\
S & = &	e^2\int\limits_{-\infty}^\infty\frac{d\omega}{2\pi} T(\omega)\big[ 1 - T(\omega) \big],
	\label{ScatteringResults}
\end{eqnarray}
where $e$ is the elementary charge (chosen negative).
The time-dependent occupation probability $n_d(t)$
can be expressed with Green's functions
\cite{Cini80, Stefanucci04}.
\new{9.A}{We use that at infinite bias the left Fermi function is unity
	and the right is zero.}
For simplicity we assume $n_d(0)=0$ and thus \new{9.A}{
\begin{eqnarray}
n_d(t)
& = &	-i\sum_{k} G^R_{d,kL}(t) i  G^A_{kL,d}(-t)	\nonumber\\
& = &	\sum_{k} |G^R_{d,kL}(t)|^2,
\end{eqnarray}
}
where we have used
\mbox{$G^A_{ka,d}(-t)=\left[G^R_{d,ka}(t)\right]^*$}.

The required Green's functions are obtained
using equations of motion
\cite{Mahan00, Hewson93, Haug08, Bruus04}
and read
\begin{equation}
G_{d,kL}^R(t)=t_{kL}\int\limits_{-\infty}^\infty\frac{d\omega}{2\pi}
	\frac{1}{\omega-\epsilon_{kL}+i0^+}
	\frac{e^{-i\omega t}}{\omega-\epsilon_d-\Sigma^R(\omega)},
\end{equation}
which, upon inserting
\mbox{$1=\int_{-\infty}^\infty\delta(\omega-\epsilon_{kL})d\omega$},
yields the explicit result
\begin{eqnarray}
n_d(t)
& = &	\int\limits_{-\infty}^\infty\frac{d\omega}{2\pi}
	\Gamma_L\times 	\nonumber \\
&&	\left|\int\limits_{-\infty}^\infty\frac{d\omega'}{2\pi}
        \frac{1}{\omega'-\omega+i0^+}
	\frac{e^{-i\omega' t}}
	{\omega'-\epsilon_d-\Sigma^R(\omega')}\right|^2.	\nonumber \\
&&	\label{n}
\end{eqnarray}
This can alternatively be obtained from a direct calculation
without Green's functions \cite{Schaller08b}.

For the Lorentzian-shaped right tunneling rate (\ref{Gamma}),
the Kramers-Kronig-relation (\ref{KramersKronig}) yields the level-shift function
\begin{equation}
\Lambda(\omega)
  = \frac{\Gamma_{R,0}^2}{2\pi}
    \frac{\omega-\varepsilon_R}{(\omega-\varepsilon_R)^2 + \delta_R^2},
\label{Lambda}
\end{equation}
and the self energy (\ref{SelfEnergy}) can be simplified to
\begin{equation}
\Sigma^{R/A}(\omega) = \mp i\frac{\Gamma_L}{2}
  + \frac{\Gamma_{R,0}^2}{2\pi}\frac{1}{\omega-\varepsilon_R\pm i\delta_R}.
\end{equation}
With Lorentzian shaped tunneling rates we can analytically
integrate the expressions for current and noise, (\ref{ScatteringResults}),
and with the abbreviations
\mbox{$\bar\Gamma:=\Gamma_L+2\delta_R$}
and \mbox{$\epsilon:=\epsilon_d-\varepsilon_R$}, we obtain
\begin{eqnarray}
I & = &	e\frac{2\Gamma_L\delta_R\bar\Gamma\ \Gamma_{R,0}^2/\pi}{
	  (\Gamma_L\delta_R+\Gamma_{R,0}^2/\pi)\bar\Gamma^2+4\epsilon^2\Gamma_L\delta_R}\ ,
	\nonumber	\\
S & = &	Ie\Bigg( 1 - \frac{2\Gamma_L\delta_R\Gamma_{R,0}^2}{\pi\bar\Gamma}\times	\nonumber	\\
&&	\frac{
	  4\epsilon^2(\Gamma_L^3+8\delta_R^3)+\bar\Gamma^3(\bar\Gamma^2+2\Gamma_L\delta_R+2\Gamma_{R,0}^2/\pi)}{
	  \big[(\Gamma_L\delta_R+\Gamma_{R,0}^2/\pi)\bar\Gamma^2+4\epsilon^2\Gamma_L\delta_R\big]^2
	}\Bigg).	\nonumber\\
&&	\label{ExplicitResults}
\end{eqnarray}

\subsection{Comparison with double dot model}
\begin{figure*}
\includegraphics[width=\linewidth]{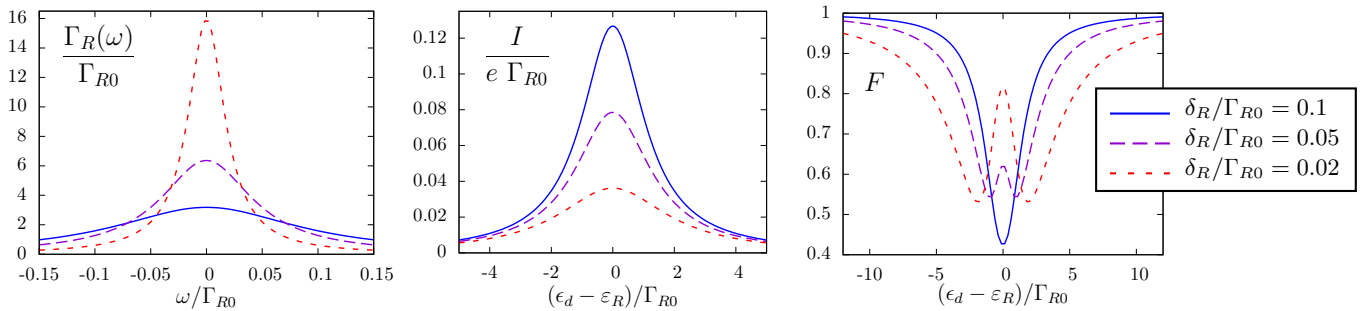}
\caption{\label{GJF}
	The left plot shows the right tunneling rate $\Gamma_R(\omega)$
	as a function of the energy $\omega$.
	We further show how the current $I$ and Fano factor $F$
	behave as a function of the detuning $\epsilon_d-\varepsilon_R$.
	The bandwidth $\delta_R$ takes the three values
	shown on the right.
	For all curves we have $\Gamma_L=\Gamma_{R,0}$.
	Results from exact Green's functions.}
\end{figure*}
Our model has an exact correspondence with
an effective \new{9c}{wide band} two level Fano-Anderson model
\cite{Elattari00a}
with Hamiltonian
\begin{eqnarray}
\bar H & = &	\bar\epsilon_L d^\dagger_L d_L
	+ \bar\epsilon_R d^\dagger_R d_R
	+ \bar T_C d^\dagger_L d_R + \bar T_C^* d^\dagger_R d_L	\nonumber	\\
&&	+ \sum_{k,a} \bar\epsilon_{ka} c^\dagger_{ka} c_{ka}
	+ \sum_{k,a} ( \bar t_{ka} d_a^\dagger c_{ka} + \mathrm{h.c.} )
	\label{H2}
\end{eqnarray}
with a left and a right dot state,
$d_L$ and $d_R$, and a coupling $\bar T_C$ between them.
Each lead couples only to its adjacent dot.
The transmission probability is \cite{Sumetskii91}
\begin{equation}
\bar T(\omega)
= \frac{\bar\Gamma_L\bar\Gamma_R |\bar T_C|^2}{
  \big|(\omega-\bar\epsilon_L+i\bar\Gamma_L/2)(\omega-\bar\epsilon_R+i\bar\Gamma_R/2)
  -|\bar T_C|^2 \big|^2}.
\end{equation}
Comparison to Eq. (\ref{SingleDotT}),
with appropriate $\Gamma_R(\omega)$
and $\Lambda(\omega)$,
reveals the exact mapping
\mbox{$\bar T(\omega)=T(\omega)$}
by the correspondence
\begin{eqnarray}
\bar\Gamma_L	& \leftrightarrow &	\Gamma_L,	\nonumber\\
|\bar T_C|^2	& \leftrightarrow &	\Gamma_{R,0}^2/(2\pi),	\nonumber\\
\bar\Gamma_R	& \leftrightarrow &	2\delta_R,	\nonumber\\
\bar\epsilon_L	& \leftrightarrow &	\epsilon_d,	\nonumber\\
\bar\epsilon_R	& \leftrightarrow &	\varepsilon_R.
\label{ExactMapping}
\end{eqnarray}
Since the cumulant generating function can be expressed
solely in terms of the transmission probabilities\cite{BLA00, Levitov93, Lee95},
not only the first two cumulants \cite{Elattari02},
but \emph{all} current cumulants, of the two models coincide.

When we plot current and Fano factor $F=S/(Ie)$ as a function
of the detuning $\epsilon_d-\varepsilon_R$ in Fig. \ref{GJF}
we find the typical structures of the corresponding quantities
for the non-interacting double quantum dot\cite{Elattari02, Kiesslich06}:
The current exhibits a maximum when the detuning vanishes.
For the noise we find a minimum in resonance
for broad tunneling rates, when $\delta_R$ becomes
sufficiently small a new maximum in the Fano factor
at resonance grows out of the minimum.

\section{Non-Markovian dynamics}

\subsection{Non-Markovian Master Equation (NMME)}
To treat non-Markovian effects in a transport master equation framework
one writes the master equation as an
integro-differential equation
for the $n$-resolved reduced density matrix
\begin{equation}
\dot\rho_n(t)=\sum_{n'}\int\limits_0^t \mathcal{W}_{n-n'}(t-t')\rho_{n'}(t')dt',
\label{BornStart}
\end{equation}
where $n$ denotes the number of charges
that have crossed the considered system.
\new{3 and 7}{
In Appendix \ref{AppA} we show how our system
can be described with such an equation using the Born approximation.
All \mbox{$\mathcal{W}_n(t)$} vanish except
when \mbox{$n=0$} or \mbox{$n=1$}, where we find
\begin{eqnarray}
\mathcal{W}_0(t)
& = &	\left(\begin{array}{cc}
	  -\gamma_L(t) & 0 \\ \gamma_L(t) & -\gamma_R(t)
	\end{array}\right),	\nonumber\\
\mathcal{W}_1(t)
& = &	\left(\begin{array}{cc}
	  0 & \gamma_R(t) \\ 0 & 0
	\end{array}\right)
\end{eqnarray}
with
\begin{eqnarray}
\gamma_R(t)
& = &	2\int\frac{d\omega}{2\pi}\Gamma_R(\omega)
	 \cos[(\omega-\epsilon_d)t]	\nonumber\\
& = &	\frac{\Gamma_{R,0}^2}{\pi} e^{-\delta_Rt}
	 \cos[(\varepsilon_R-\epsilon_d)t],	\nonumber\\
\gamma_L(t)
& = &	2\int\frac{d\omega}{2\pi}\Gamma_L\cos[(\omega-\epsilon_d)t]
	  \nonumber\\
& = &	2\Gamma_L\delta(t).
\end{eqnarray}
We proceed with performing a Fourier summation and a Laplace transform via
\begin{eqnarray}
\hat{\mathcal{W}}(\chi,z)
& \equiv &	\sum_ne^{in\chi}\int\limits_{0}^\infty e^{-zt} \mathcal{W}_n(t) dt	\nonumber\\
& = &	\hat{\mathcal{W}}_0(z) + \hat{\mathcal{W}}_1(z) e^{i\chi}	\nonumber\\
& = &	\left(\begin{array}{cc}
	-\hat\gamma_L(z) & e^{i\chi}\hat\gamma_R(z) \\
	\hat\gamma_L(z) & -\hat\gamma_R(z)
	\end{array}\right),
	\label{LaplaceKernel}
\end{eqnarray}
where
\begin{eqnarray}
\hat\gamma_R(z)
& = &	\frac{\Gamma_{R,0}^2}{\pi}
	 \frac{z+\delta_R}{(z+\delta_R)^2+(\varepsilon_R-\epsilon_d)^2},
	 \nonumber\\
\hat\gamma_L(z)
& = &	\Gamma_L.
\end{eqnarray}
We will use the abbreviation
\mbox{$\hat\gamma(z)=\hat\gamma_L(z)+\hat\gamma_R(z)$}.
In order to avoid the tedious inverse Laplace transform
we can use a recently developed elegant method
to evaluate current and noise \cite{Flindt07,FLI08}.
As it has only applied a few times until now
\cite{Flindt07, FLI08, Flindt08, Urban08}
we show the explicit calculation for our model
in appendix \ref{App1} and
obtain the following formulae:
\begin{eqnarray}
I
& = &	e\frac{\hat\gamma_L(0)\hat\gamma_R(0)}{\hat\gamma(0)},
	\nonumber	\\
S
& = &	Ie
	\Big\{\frac{\hat\gamma_R^2(0)}
	{\hat\gamma^2(0)}\big[1+2\hat\gamma_L'(0)\big]
	\label{MENoise}	\\
&& \quad
	+\frac{\hat\gamma_L^2(0)}
	{\hat\gamma^2(0)}\big[1+2\hat\gamma_R'(0)\big]\Big\}.
	\nonumber
\end{eqnarray}
}

\new{3 and 7}{
We easily obtain the occupation of the dot
by evaluating the density matrix in Laplace space,
\mbox{$\hat\rho(z)=[z-\hat{\mathcal{W}}(z)]^{-1}\rho(t=0)$}.
To obtain the time-resolved dynamics, one has to perform
the inverse Laplace transform (Bromwich integral)
by collecting all the corresponding residues.
}

\new{3 and 7}{\emph{Markovian master equation (MME)} ---
A Markovian master equation follows from (\ref{BornStart})
by the integration
\mbox{$\mathcal{L}_n=\int_0^\infty\mathcal{W}_n(t)dt
= \hat{\mathcal{W}}_n(z=0)$}
and leads to
\begin{equation}
\mathcal{L}_0
= \left(\begin{array}{cc}-\Gamma_L & 0 \\ \Gamma_L & -\Gamma_R(\epsilon_d)
  \end{array}\right),\quad
\mathcal{L}_1
= \left(\begin{array}{cc}0 & \Gamma_R(\epsilon_d) \\ 0 & 0
  \end{array}\right).
\end{equation}
}
When we use this to evaluate noise we end up
with the same result as in Eq. (\ref{MENoise}),
but without the derivatives \mbox{$\hat\gamma_L'(0)$}
and \mbox{$\hat\gamma_R'(0)$}.
For the equilibrium density matrix we find
\mbox{$\rho_{00}(t=\infty)=\Gamma_R(\epsilon_d)/
  \big(\Gamma_L+\Gamma_R(\epsilon_d)\big)$},
which is identical with the NMME result.
The advantage of the additional Markov approximation
is that the positivity of the density matrix will be conserved
as one obtains a Lindblad type master equation \cite{Lindblad76}.
The disadvantage is that some information about the
shape of the tunneling rates is lost.

\subsection{Dynamical Coarse Graining (DCG)}
A second approach to quantum transport is the recently
developed dynamical coarse graining method \cite{Schaller08}.
The coarse graining method
\new{8.A}{is also a second order weak coupling approximation
  although it can be extended to higher orders.}
Instead of solving a single master equation,
it solves a continuous set
\mbox{$\dot\rho^\tau(t)=\mathcal{L}^\tau\rho^\tau(t)$}
and then interpolates through the solutions 
\mbox{$\rho^\tau(t)=e^{\mathcal{L}^\tau\cdot t}\rho_0$} at \mbox{$t=\tau$}.
The coarse grained Liouvillian can be derived \cite{Schaller08} by matching the second order
expansion of the formal solution in the interaction picture
$\tilde\chi(t) = \tilde U(t)\chi(0)\tilde U^\dagger(t)$ (where
$\tilde U(t) = T \exp\left\{-i\int_0^t H_{SB}(t')dt'\right\}$
with the time ordering operator $T$) with the second order expansion
of \mbox{$\rho^\tau(t)=e^{\mathcal{L}^\tau t}\rho^\tau(0)$}
at time $t=\tau$.
For our specific model,
\new{8A}{
	following reference \cite{Schaller08b} we obtain
}
\begin{equation}
\mathcal{L}^\tau = \int\limits_{-\infty}^\infty \frac{d\omega}{2\pi}
  \tau\ \mathrm{sinc}^2\frac{(\omega-\epsilon_d)\tau}{2}\left(
  \begin{array}{cc}-\Gamma_L(\omega) & \Gamma_R(\omega)\\
    \Gamma_L(\omega) & -\Gamma_R(\omega) \end{array}\right),
\end{equation}
\new{8B}{where \mbox{$\mathrm{sinc}\ x\equiv\frac{\sin x}{x}$}.}
The Coarse Graining method combines the two advantages
of the Born and the Born-Markov approximation:
For finite times it is sensitive to the shape of the tunneling rates,
and at the same time it preserves positivity
since the $\mathcal{L}^\tau$ are of Lindblad form.
Due to the identity
\mbox{$\lim_{\tau\to\infty} \tau\cdot\mathrm{sinc}^2\left[\frac{(\omega-\epsilon_d)\tau}{2}\right]
 = 2\pi\delta(\omega-\epsilon_d)$}
we see that for large times the coarse graining method yields
the same steady state density matrix as the two other approximations.
Also for current and Fano factor we have reproduced the results of the MME.
\new{5}{In contrast to fixed graining-time derivations of master equations \cite{KNE08,Lidar01}, 
DCG dynamically adapts the coarse-graining time with the physical time, which in the 
long-time limit yields the Born Markov secular approximation. Both approaches yield
completely positive maps but may lead to different stationary states.}

\section{Discussion}
\subsection{Current}
\begin{figure}
\includegraphics[width=\linewidth]{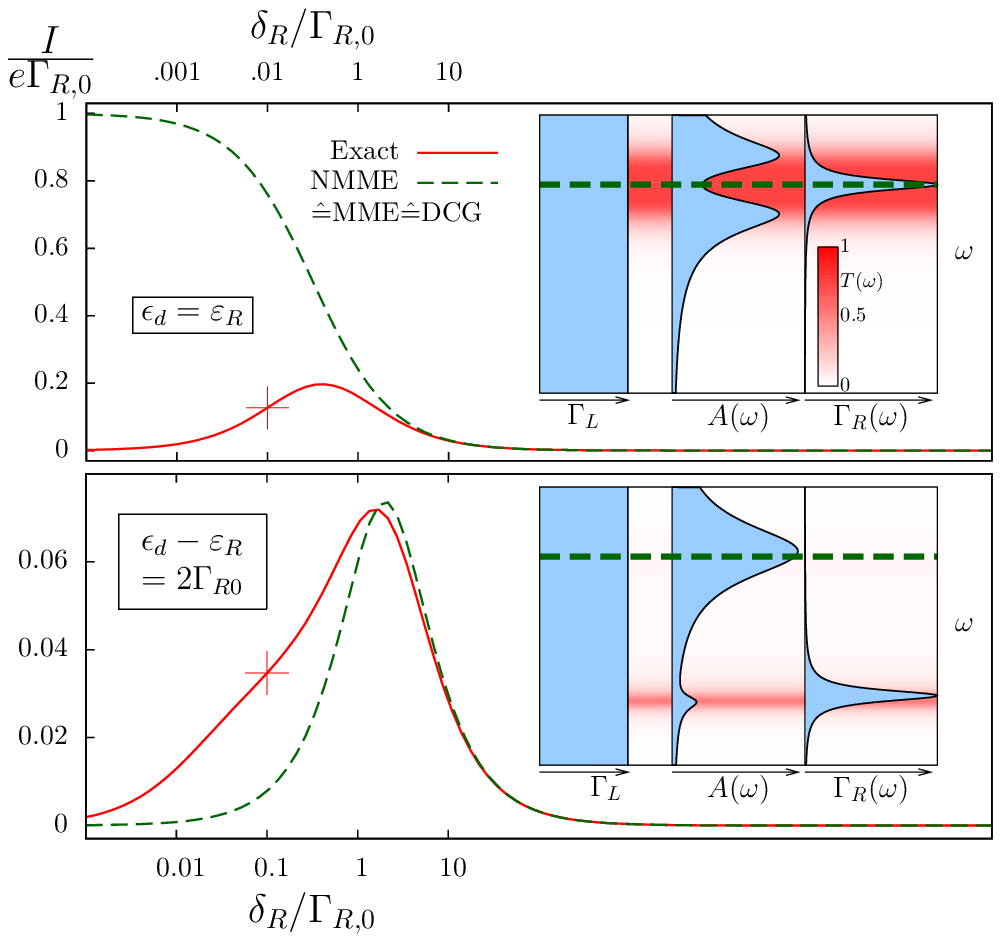}
\caption{\label{IT}
	Current $I$ as a function of the width $\delta_R$
	on a logarithmic scale
	in absence (upper part) and
	presence (lower part) of detuning,
	exact (red, solid) and with the NMME$\hat=$MME$\hat=$DCG (green, dashed)
	with the parameters \mbox{$\Gamma_L=\Gamma_{R,0}$}.\\
	Inset: For the marked points at \mbox{$\delta_R/\Gamma_{R,0}=0.1$}
	we plot as functions of $\omega$ the left tunneling rate $\Gamma_L$,
	the dot spectral function $A(\omega)$,
	and the right tunneling rate \mbox{$\Gamma_R(\omega)$}, with
	a red density plot
	of the transmission probability $T(\omega)$ in the background.
	The dashed green line marks the level of the dot state.
	The master equations depend only on $\Gamma_R(\epsilon_d)$,
	while the exact solution is sensitive
	to the detailed shape of \mbox{$A(\omega)$} and \mbox{$T(\omega)$}.}
\end{figure}
To get a first idea of the difference between the exact solution
and the NMME it is instructive to take a look at the
stationary current $I$.
The stationary current is not sensitive to non-Markovian effects \cite{FLI08},
since it only depends on the steady state occupation,
and thus all three presented approximations coincide.
The two plots in Fig. \ref{IT}
show the current as a function
of the width of the right tunneling rate, $\delta_R$.
We observe that for large $\delta_R$ (Markovian limit)
the master equations meet the exact solution very well,
while for small $\delta_R$ enormous deviations occur.
In the absence of detuning,
\mbox{$\epsilon_d-\varepsilon_R=0$},
the master equations overestimate the current,
but they underestimate it for sufficiently large detuning.
As an inset in Fig. \ref{IT} we choose the value $\delta_R/\Gamma_{R,0}=0.1$
and show how the spectral function of the dot state $A(\omega)$,
the right tunneling rate $\Gamma_R(\omega)$
and the transmission coefficient $T(\omega)$ behave there.

Without detuning the spectral function exhibits a minimum
at energies where the tunneling rate is maximal.
Thus the true quantum mechanical transmission is smaller than
the master equation result, which depends exclusively on the value of the
tunneling rate at the dot level, $\Gamma_R(\epsilon_d)$.
This leads to the underestimation of the current by the master equations
when detuning is present,
because $\Gamma_R(\epsilon_d)$ is very small then.
The exact result, by contrast,
takes into account that the spectral function exhibits
a small side peak near the maximum of $\Gamma_R(\omega)$
that enhances the probability for tunneling.

\subsection{Fano factor}
\begin{figure}
\includegraphics[width=\linewidth]{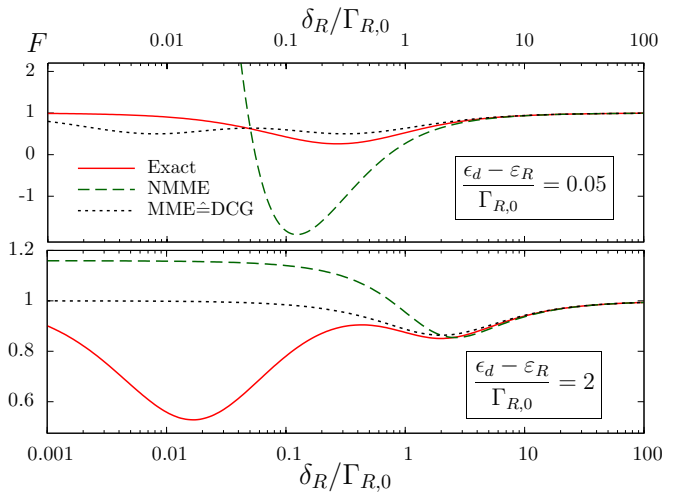}
\caption{\label{FF}
	Fano factor $F$ as a function of the
	width $\delta_R$
	on a logarithmic scale.
	Parameters are
	\mbox{$\Gamma_L=\Gamma_{R,0}$}
	and the detunings
	\mbox{$\epsilon_d-\varepsilon_R$}
	indicated in each plot.}
\end{figure}
In Fig. \ref{FF} we show results for the Fano factor \mbox{$F=S/(Ie)$}
(exact, with NMME and with MME/DCG).
When the condition
\begin{equation}
\frac{\delta_R}{\Gamma_{R,0}}\gg 1
\end{equation}
is fulfilled, all formalisms agree.
Furthermore, for constant tunneling rates,
the NMME yields the Markovian result
\mbox{$F=\frac{\Gamma_L^2+\Gamma_R^2(\epsilon_d)}{[\Gamma_L+\Gamma_R(\epsilon_d)]^2}$}.

The exact Fano factor shows one minimum for small
detuning and two minima for large detuning,
which originates from the double peak structure of
the spectral function shown in Fig. \ref{IT}.
However, there is no simple quantitative connection
between the locations of the extrema in the Fano factor
and in the spectral function.
This is similar to the Fano factor as function of the
detuning in Fig. \ref{GJF}, where for small width
$\delta_R$ a second minimum appears, that is again
not connected to the spectral function's properties
in a simple way.

The NMME reproduces one of the two \new{9B}{minima}, but not both,
and it produces super-Poissonian noise, where it should not appear.
If the detuning $\epsilon_d-\varepsilon_R$ becomes very small,
the NMME even overestimates the minimum so strongly
that it yields an unphysical negative Fano factor.

The MME is not everywhere close to the exact result,
but it is, on average, closer than the NMME result,
and per construction yields physical results (Lindblad form).
For the DCG method we have found that it
yields the same result as the MME.

\subsection{Time resolved occupation probabilities}
\begin{figure}
\includegraphics[width=\linewidth]{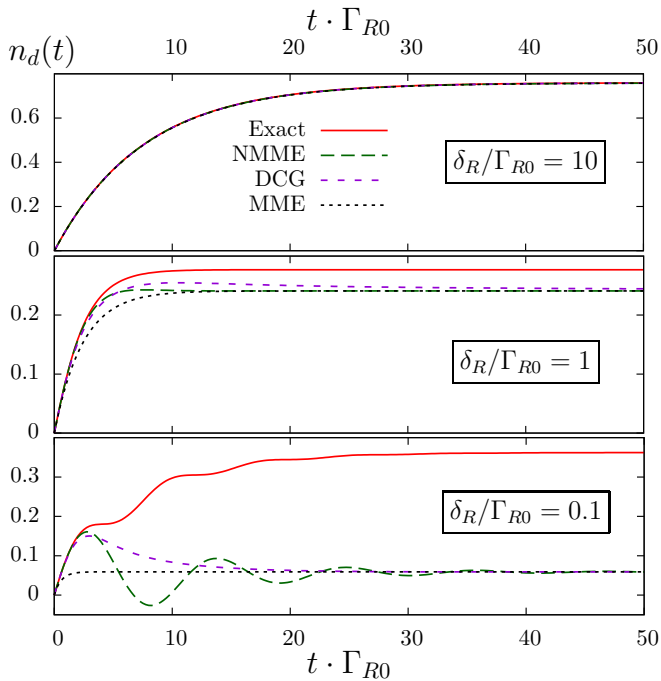}
\caption{\label{TimeDependentN}
	Time dependent occupation probability $n_d(t)$,
	exact solution and solution in the three presented approximations
	Parameters are \mbox{$\epsilon_d=\Gamma_{R,0}$} and
	\mbox{$\epsilon_d-\epsilon_R=\Gamma_L=0.1\Gamma_{R,0}$}.
	The width $\delta_R$ takes the values denoted in each plot.}
\end{figure}
\begin{figure}
\includegraphics[width=\linewidth]{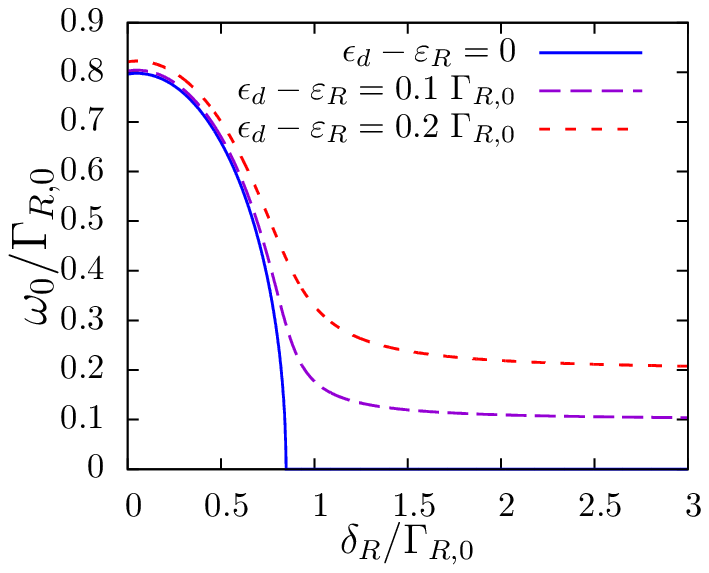}
\caption{\label{w0}
	Frequency $\omega_0$ as a function of
	the detuning \mbox{$\omega_d-\omega_R$}
	for the parameters
	\mbox{$\Gamma_L=0.1\Gamma_{R,0}$}
	and
	\mbox{$\epsilon_d=\Gamma_{R,0}$}.
	}
\end{figure}
One key advantage of the DCG approach
results from its ability to preserve positivity
and at the same time to be more sensitive to the shape of the tunneling rates 
than the MME \cite{Schaller08}.
We therefore evaluate the time dependent occupation probability
of the localized level
with four different methods: exact,
with the NMME,
with the MME,
and with DCG.
In Fig. \ref{TimeDependentN} we find
that at sufficiently wide bands, i.e., $\delta_R\ge\Gamma_{R,0},$
all three approximations meet the exact result very well.

For smaller $\delta_R$ one recognizes that in the stationary limit,
the three approximations coincide with each other
but not with the exact solution.
The most prominent feature of the exact occupation
\mbox{$n_d(t)$}
are oscillations as a function of time $t$.
The NMME approximation captures the oscillations as
non-Markovian features of the reservoirs,
as it should, but it strongly overestimates them.
If the steady state is sufficiently close to zero,
the NMME can lead to negative occupation probabilities,
like in the third plot.

In order to better understand the exact result,
we explicitly perform the two integrations
in Eq. (\ref{n}), which leads to
\begin{equation}
\begin{array}{l}
\displaystyle
	n_d(t)\ =	\\
\displaystyle
	\Gamma_L\frac	{\delta_R(\bar\Gamma^2+4\epsilon^2)+\bar\Gamma\ \Gamma_{R,0}^2/\pi}
			{\delta_R(\bar\Gamma^2+4\epsilon^2)+\bar\Gamma^2\Gamma_{R,0}^2/\pi}
		\\
\displaystyle
	+i\Gamma_L\frac	{(\omega_1-\epsilon_R+i\delta_R)(\omega^*_1-\epsilon_R-i\delta_R)}
			{(\omega_1-\omega_2)(\omega_1^*-\omega_2^*)(\omega_1-\omega_1^*)}
		e^{i(\omega_1^*-\omega_1)t}
		\\
\displaystyle
	+i\Gamma_L\frac	{(\omega_2-\epsilon_R+i\delta_R)(\omega_1^*-\epsilon_R-i\delta_R)}
			{(\omega_1-\omega_2)(\omega_1^*-\omega_2^*)(\omega_2-\omega_2^*)}
		e^{i(\omega_2^*-\omega_2)t}
		\\
\displaystyle
	+2\Gamma_L\mathrm{Im}\left(\frac
		{(\omega_1^*-\epsilon_R-i\delta_R)(\omega_2-\epsilon_R+i\delta_R)}
		{(\omega_1-\omega_2)(\omega_1^*-\omega_2^*)(\omega_2-\omega_1^*)}
		e^{(i\omega_0-\bar\Gamma/2) t} \right),
\end{array}
\label{nResult}
\end{equation}
with the abbreviations
\mbox{$\bar\Gamma:=\Gamma_L+2\delta_R$}
and \mbox{$\epsilon:=\epsilon_d-\epsilon_R$}.
The frequencies $\omega_1$ and $\omega_2$
are the poles of the spectral function $A(\omega)$
with negative imaginary part,
and $\omega_1^*$ and $\omega_2^*$ are its other two poles.
The first part of \mbox{$n(t)$} in Eq. (\ref{nResult})
corresponds to the steady state.
The next two lines describe how the system
performs exponential decay towards this steady state,
because the exponents are real and negative.
The last line of Eq. (\ref{nResult}) is responsible
for the oscillations, because its exponent
contains the imaginary part \mbox{$i\omega_0 t$},
which is given by
\begin{eqnarray}
\omega_0
& = &	\mathrm{Re}\Bigg\{\bigg[
	(\epsilon_d+\varepsilon_R-i\bar\Gamma/2)^2	\nonumber\\
&&	-4(\varepsilon_R-i\delta_R)(\epsilon_d-i\Gamma_L/2)
	+4\frac{\Gamma_{R,0}^2}{2\pi}\bigg]^{1/2}\Bigg\}.
	\label{w0Frequenz}
\end{eqnarray}
If we switch off the coupling to the left side
and the width on the right, i.e.,
\mbox{$\Gamma_L\to 0$} and \mbox{$\delta_R\to 0$},
we recover the frequency:
\mbox{$\omega_0=\sqrt{\epsilon^2+4\frac{\Gamma_{R,0}^2}{2\pi}}$}
of coherent oscillations in an isolated two-level system
(remember the mapping to the double dot model in Eq. (\ref{ExactMapping})).
In Fig. \ref{w0} one recognizes that in the absence of the detuning
\mbox{$\epsilon_d-\varepsilon_R$},
the frequency $\omega_0$
completely vanishes with a nonanalyticity
where \mbox{$\delta_R/\Gamma_{R,0}$} is of the order of $1$.
This nonanalyticity appears,
when the radicand in Eq. (\ref{w0Frequenz})
changes sign (at \mbox{$\epsilon_d=\varepsilon_R$}  it is purely real).
When detuning is present,
the frequency $\omega_0$
does not completely vanish for \mbox{$\delta_R\to\infty$},
but the oscillations of \mbox{$n_d(t)$} are damped with the rate
\mbox{$\bar\Gamma/2=\delta_R+\Gamma_L/2$},
such that in the Markovian limit,
\mbox{$\delta_R\to\infty$},
no oscillations survive.

\section{Conclusions}
The main message of this paper is that one must be careful
with non-Markovian master equations (NMMEs).
We know that what we have here called ``Markovian master equations'' (MMEs)
may lead to incorrect results when non-Markovian effects are strong.
However, we have found that in the non-Markovian regime
NMMEs can be worse than MMEs.
On the one hand, NMMEs can give quantitative errors
such as occur in Fig. (\ref{FF}).
On the other, the positivity of the density matrix is not in general conserved,
which can lead to unphysical results like a negative Fano factor
and occupation probability, as in Fig. (\ref{FF}) and (\ref{TimeDependentN}).
We emphasize that the failure of the NMME does not result
from the new evaluation techniques we have used \cite{FLI08},
but instead from the way the Born approximation is performed
in the derivation of Eq. (\ref{BornStart}).
We do not want to \new{9c}{discourage using} of NMMEs in general,
but each time they are applied one should reason carefully
why one expects them to yield better results than MMEs.

We have tried the dynamical coarse graining (DCG) method as an alternative
approach that is more sophisticated than the MME, but still
in Lindblad form.
For the time-independent quantities $I$, $F$ and \mbox{$n(t\to\infty)$}
we have found no improvement in comparison to the MME.
For the time-dependent occupation probability however,
DCG does yield better results, at least for small times.
We expect a similar improvement for the frequency-dependent Fano factor.

For the model considered here, the condition
\mbox{$\frac{\Gamma_{R,0}}{\delta_R}\ll 1$}
must be fulfilled to ensure a good quality of the three
presented weak coupling approximations.
\new{4}{
It is certainly interesting to explore higher-order corrections to the presented
perturbative approximations. However, such extensions should be treated with caution, 
since nice features such as positivity may be lost when higher orders are included.}

\begin{acknowledgments}
We would like to thank
S. Gurvitz,
C. L\'opez-Mon\'is,
U. Kleinekath\"ofer,
and
T. Novotn\'y
for helpful discussions.
\new{12}{This work was supported by DFG grant BR 1528/5-1.}
\end{acknowledgments}

\bibliography{bib,references}


\appendix

\section{Derivation of the kernel}
\label{AppA}
\new{3 and 7}{
We decompose our Hamiltonian into the free part
\mbox{$H_0=\epsilon_dd^\dagger d+\sum_{k,a}\epsilon_{ka}c^\dagger_{ka}c_{ka}$}
and the coupling
\mbox{$V=\sum_{k,a}(t_{ka}d^\dagger c_{ka}+t^*_{ka}c^\dagger_{ka}d)$}.
This enables us to define the density matrix in the interaction picture,
\mbox{$\tilde\chi(t)$}.
It obeys the Liouville-von-Neumann equation
\begin{equation}
\dot{\tilde\chi}(t) = -i[\tilde V(t),\tilde\chi(t)],
\label{AppLvN}
\end{equation}
where \mbox{$\tilde V(t)=e^{iH_0t}Ve^{-iH_0t}$}
is the interaction picture version of the coupling.
A tilde will mark the interaction picture in all following text.
By iterating Eq. (\ref{AppLvN}) twice, we obtain
\begin{equation}
\dot{\tilde\chi}(t)=-i[\tilde V(t),\tilde\chi(0)]
	-\int\limits_0^tdt'[\tilde V(t),[\tilde V(t'),\tilde\chi(t')]].
	\label{AppLvNIterate}
\end{equation}
At this point we perform the second-order weak coupling approximation
by replacing the full density matrix \mbox{$\tilde\chi(t')$}
by a tensor product of the reduced density matrix \mbox{$\tilde\rho(t)$}
and the bath density matrix $R_0$,
which we assume to be constant in time (Born approximation).
The partial trace over the
first commutator in equation (\ref{AppLvNIterate}) vanishes.
To proceed we resolve the coupling into system operators $S_i$
and bath operators $B_i$ such that
\begin{equation}
\tilde V(t) = \tilde S_1(t)
	\tilde B_1(t)
	+\tilde B_2(t)
	\tilde S_2(t).
\end{equation}
where
\begin{equation}
\begin{array}{rclcrcl}
\tilde S_1(t)
& = &	\tilde d^\dagger(t),
& \qquad\qquad &
\tilde B_1(t)
& = &	\displaystyle\sum_{k,a}t_{ka}\tilde c_{ka}(t),
	\\[3mm]
\tilde S_2(t)
& = &	\tilde d(t),
& \qquad &
\tilde B_2(t)
& = &	\displaystyle\sum_{k,a}t^*_{ka}\tilde c^\dagger_{ka}(t).
\end{array}
\end{equation}
In the calculation that follows we find that the off-diagonal elements
of \mbox{$\tilde\rho(t)$} decouple from the diagonal ones.
With this knowledge we can choose the off-diagonals to be zero
and neglect them in the density matrix,
which we can therefore consider as a vector with two entries:
\begin{equation}
\tilde\rho(t) = \tilde\rho_{00}(t)dd^\dagger+\tilde\rho_{11}(t)d^\dagger d
	= \left(\begin{array}{c}
		\tilde\rho_{00}(t) \\ \tilde\rho_{11}(t)
	\end{array}\right).
\end{equation}
To evaluate the double commutator in equation (\ref{AppLvNIterate})
is lengthy but straight forward and yields
\begin{equation}
\dot{\tilde\rho}(t) = \int\limits_0^t dt'\mathcal{W}(t-t')\tilde\rho(t')
\end{equation}
with
\begin{equation}
\mathcal{W}(t) = \left(\begin{array}{cc}-\gamma_L(t) & \gamma_R(t) \\ \gamma_L(t) & -\gamma_R(t) \end{array}\right).
\end{equation}
For the entries of \mbox{$\mathcal{W}(t)$} we need to explicitly
perform traces over the reservoirs:
\begin{eqnarray}
\gamma_L(t)
& = &	e^{i\epsilon_dt}\mathrm{Tr}
		\big\{\tilde B_2(0)\tilde B_1(t)R_0\big\}
	\nonumber\\
&&	+ e^{-i\epsilon_dt}\mathrm{Tr}
		\big\{\tilde B_2(t)\tilde B_1(0)R_0\big\}.
\end{eqnarray}
With infinite bias, i.e., \mbox{$R_0=\sum_kc^\dagger_{kL}c_{kL}$}
this becomes
\begin{eqnarray}
\gamma_L(t)
& = &	\sum_k|t_{kL}|^2\cdot 2\cos[(\epsilon_{kL}-\epsilon_d)t]
	\nonumber \\
& = &	2\int\limits_{-\infty}^\infty\frac{d\omega}{2\pi}\Gamma_L(\omega)
	 \cos[(\omega-\epsilon_d)t].
\end{eqnarray}
In the same manner we get
\begin{equation}
\gamma_R(t)\ = \ 2\int\limits_{-\infty}^\infty\frac{d\omega}{2\pi}\Gamma_R(\omega)
	\cos[(\omega-\epsilon_d)t].
\end{equation}
Physically, the upper right matrix element of
\mbox{$\mathcal{W}(t)$}
describes a jump from the dot to the right lead.
Thus we can say that it increases the number
of passed electrons by one,
while all other matrix elements leave it unchanged.
Thus we distinguish
\begin{eqnarray}
\mathcal{W}_0(t)
& = &	\left(\begin{array}{cc}
	  -\gamma_L(t) & 0 \\ \gamma_L(t) & -\gamma_R(t)
	\end{array}\right),	\\
\mathcal{W}_1(t)
& = &	\left(\begin{array}{cc}
	  0 & \gamma_R(t) \\ 0 & 0
	\end{array}\right).
\end{eqnarray}
}

\section{Evaluation of current and noise}
\label{App1}
We use a bra-ket like notation where the equilibrium state
is represented by the ket
\begin{equation}
|0\rrangle
= \lim_{t\to\infty}\rho(t)
= \frac{1}{\hat\gamma(0)}\left(\begin{array}{c}
	\hat\gamma_R(0)	\\
	\hat\gamma_L(0)
\end{array}\right),
\end{equation}
that we obtain by setting \mbox{$\dot\rho=0$} in Eq. (\ref{BornStart}).
We define the bra
\begin{equation}
\llangle\tilde 0|=(1,1),
\end{equation}
and construct the projector
\begin{equation}
\mathcal{P}
= 	\mathcal{P}^2 =	|0\rrangle\llangle\tilde0|
=	\frac{1}{\hat\gamma(0)}\left(\begin{array}{cc}
	\hat\gamma_R(0) & \hat\gamma_R(0)	\\
	\hat\gamma_L(0) & \hat\gamma_L(0)
	\end{array}\right),	\\
\end{equation}
and the projector
\mbox{$\mathcal{Q}=\mathcal{Q}^2=\hat1-\mathcal{P}$}.
The resolvent of 
\new{3 and 7}{
	the Laplace-transformed kernel from Eq. (\ref{LaplaceKernel})
}
is
\mbox{$\mathcal{R}(\Lambda,\chi,z)
	:=\mathcal{Q}[\mathcal{W}(\chi,z)-\Lambda\cdot\mathbbm{1}]^{-1}\mathcal{Q}$}.
We only need its value at zero,
\mbox{$\mathcal{R} := \mathcal{R}(0,0,0)$}, which is
\begin{equation}
\mathcal{R}
\ = \	\frac{1}{(\hat\gamma_L(0)+\hat\gamma_R(0))^2}
	\left(\begin{array}{cc}
	  -\hat\gamma_L(0) & \hat\gamma_R(0) \\
	  \hat\gamma_L(0) & -\hat\gamma_R(0)
	\end{array}\right).
\end{equation}
We introduce the coefficients of the kernel's Taylor series via
\begin{eqnarray}
\mathcal{W}(\chi,z)
& = & 	\bar{\mathcal{W}} + \bar{\mathcal{W}}'\chi + \dot{\bar{\mathcal{W}}} z	\nonumber\\
&&	+ \frac{1}{2}\big(\bar{\mathcal{W}}''\chi^2 + 2\dot{\bar{\mathcal{W}}}'\chi z + \ddot{\bar{\mathcal{W}}}z^2 \big)
	+ \cdots
\end{eqnarray}
The expressions for current and the zero frequency noise \cite{Flindt07, FLI08}
are now obtained as
\begin{eqnarray}
\label{NovotnyCurrent}
I & = & e\llangle\tilde0|\bar{\mathcal{W}}'|0\rrangle/i,	\nonumber\\
\label{NovotnyNoise}
S & = & e^2\Big[ \llangle\tilde0|\bar{\mathcal{W}}''|0\rrangle
		-2 \llangle\tilde0|\bar{\mathcal{W}}'\mathcal{R}\bar{\mathcal{W}}'|0\rrangle\Big]/i^2
		\nonumber\\
&&	-2i Ie\Big[\llangle\tilde0|\dot{\bar{\mathcal{W}}}'|0\rrangle
		-\llangle\tilde0|\bar{\mathcal{W}}'\mathcal{R}\dot{\bar{\mathcal{W}}}|0\rrangle\Big].
		\label{ISresults}
\end{eqnarray}
Performing the derivatives and matrix multiplications,
leads to the results of Eq. (\ref{MENoise}).

\end{document}